\documentclass[amsmath,amssymb,showpacs,twocolumn]{revtex4}
\usepackage{graphicx}% Include figure files
\makeatletter \@addtoreset{equation}{section} \makeatother
\renewcommand{\theequation}{\arabic{section}.\arabic{equation}}
\def\PT{$\mathcal{PT}$}
\def\u{\textbf{u}}
\def\PPT{$\mathcal{PPT}$}
\def\[{\begin{equation}}
\def\]{\end{equation}}

\begin{document}
\title{Symmetry breaking of solitons in two-dimensional complex potentials}
\author{Jianke Yang}
\affiliation{Department of Mathematics and Statistics, University of
Vermont, Burlington, VT 05401, USA}

\begin{abstract}
Symmetry breaking is reported for continuous families of solitons in
the nonlinear Schr\"odinger equation with a two-dimensional complex
potential. This symmetry-breaking bifurcation is forbidden in
generic complex potentials. However, for a special class of
partially parity-time-symmetric potentials, such symmetry breaking
is allowed. At the bifurcation point, two branches of asymmetric
solitons bifurcate out from the base branch of symmetry-unbroken
solitons. Stability of these solitons near the bifurcation point are
also studied, and two novel stability properties for the bifurcated
asymmetric solitons are revealed. One is that at the bifurcation
point, zero and simple imaginary linear-stability eigenvalues of
asymmetric solitons can move directly into the complex plane and
create oscillatory instability. The other is that the two bifurcated
asymmetric solitons, even though having identical powers and being
related to each other by spatial mirror reflection, can possess
different types of unstable eigenvalues and thus exhibit
non-reciprocal nonlinear evolutions under random-noise
perturbations.

\end{abstract}


\maketitle

\section{Introduction}
 Parity-time (\PT) symmetric systems are dissipative systems with
 balanced gain and loss. The name of \PT symmetry was derived from
 non-Hermitian quantum mechanics with complex potentials \cite{Bender1998}. This
 concept has since been applied to optics \cite{PT_2005,Christodoulides2007}, Bose-Einstein
 condensation \cite{BEC_PT}, electric circuits \cite{EC_PT}, mechanical systems \cite{MEC_PT} and other settings. \PT-symmetric systems have some
 remarkable properties, such as all-real linear spectra \cite{Bender1998,Ahmed2001,Musslimani2008,Zezyulin2012a} and existence
 of continuous families of solitons \cite{Musslimani2008, Wang2011, Lu2011, Hu2011,Zhu2011, Abdullaev2011, He2012, Nixon2012,
Zezyulin2012a,Dong2012,Kartashov2013,Driben2011,
Alexeeva2012,Moreira2012,Konotop2012,Barashenkov2013,Kevrekidis2013,Li2011,
Zezyulin2012b, Zezyulin2013}, which set them apart from
 other dissipative systems and make them resemble conservative
 systems. In multi-dimensions, the concept of \PT symmetry
 has been generalized to include partial parity-time (\PPT) symmetry,
 and it is shown that \PPT-symmetric systems share most of the properties of \PT systems \cite{Yang_PPT}.
 Even some non-\PT-symmetric systems have been found to posesse certain properties of \PT systems,
 such as all-real linear spectra \cite{SUSY1,SUSY2,SUSY3}  and/or existence of soliton families \cite{Tsoy2014,Konotop_OL2004}.

Symmetry-breaking bifurcation for continuous families of solitons in
symmetric systems is a fascinating phenomenon. In conservative
systems with real symmetric potentials, such symmetry breaking
occurs frequently
\cite{Weinstein_2004,Panos_2005_pitchfork,Weinstein_2008,Malomed_2008,
Sacchetti_2009,Kirr_2011,Peli_2011,Akylas_2012,Yang_classification,YangPhyd}.
That is, branches of asymmetric solitons can bifurcate out from the
base branch of symmetric solitons when the power of symmetric
solitons is above a certain threshold. But in \PT-symmetric complex
potentials, such symmetry breaking is generically forbidden
\cite{Yang_Stud14}. Mathematically the reason for this forbidden
bifurcation is that this bifurcation requires infinitely many
non-trivial conditions to be satisfied simultaneously, which is
generically impossible. Intuitively this forbidden bifurcation can
be understood as follows. Should it occur, continuous families of
asymmetric solitons would be generated. Unlike in conservative
systems, these asymmetric solitons in \PT systems would require not
only dispersion-nonlinearity balancing but also gain-loss balancing,
which is generically impossible. Surprisingly for a special class of
one-dimensional (1D) \PT-symmetric potentials of the form
$V(x)=g^2(x)+\alpha g(x)+ i g'(x)$, where $g(x)$ is a real even
function and $\alpha$ a real constant, symmetry breaking of solitons
was reported very recently \cite{Yang_breaking}. This invites a
natural question: can this symmetry breaking occur in 2D complex
potentials? If so, what type of 2D complex potentials admit such
symmetry breaking?

In this article, we study symmetry-breaking bifurcations of
continuous families of solitons in 2D complex potentials. We show
that in a special class of \PPT-symmetric separable potentials
\[  \label{Vform0}
V(x,y)=g^2(x)+\alpha g(x)+ig'(x)  + h(y),  \nonumber
\]
where $g(x)$ is a real even function, $h(y)$ an arbitrary real
function, and $\alpha$ a real constant, symmetry breaking can occur.
Specifically, from a base branch of \PPT-symmetric solitons and
above a certain power threshold, two branches of asymmetric solitons
with identical powers can bifurcate out. At the bifurcation point,
the base branch of \PPT-symmetric solitons changes stability,
analogous to conservative systems. However, the bifurcated
asymmetric solitons can exhibit new stability properties which have
no counterparts in conservative systems. One novel property is that
at the bifurcation point, the zero and simple imaginary eigenvalues
in the linear-stability spectra of asymmetric solitons can move
directly into the complex plane and create oscillatory instability.
Another novel property is that the two asymmetric solitons can
possess different types of linear-instability eigenvalues. As a
consequence, these two asymmetric solitons, which are related to
each other by spatial mirror reflection, can exhibit non-reciprocal
evolutions under random-noise perturbations.

\section{Symmetry breaking of solitons}

Nonlinear beam propagation in an optical medium with gain and loss
can be modeled by a nonlinear Schr\"odinger equation
\cite{Kivshar_book}
\begin{equation} \label{Eq:NLS}
i \Psi_z + \nabla^2 \Psi + V(x,y)\Psi + \sigma |\Psi|^2 \Psi = 0,
\end{equation}
where $z$ is the propagation distance, $(x,y)$ is the transverse
plane, $\nabla^2=\partial_{xx}+\partial_{yy}$, $V(x,y)$ is a complex
potential, and $\sigma=\pm 1$ is the sign of nonlinearity.

Solitons in Eq. (\ref{Eq:NLS}) are sought of the form
\[  \label{soliton}
\Psi(x,y,z)=e^{i\mu z}\psi(x,y),
\]
where $\mu$ is a real propagation constant, and $\psi(x,y)$ is a
localized function solving the equation
\[  \label{e:soliton}
\nabla^2 \psi + V(x,y)\psi + \sigma |\psi|^2 \psi = \mu \psi.
\]

If the complex potential $V(x,y)$ is \PT-symmetric or
\PPT-symmetric, continuous families of \PT-symmetric or
\PPT-symmetric solitons are admitted \cite{Nixon2012,Yang_PPT}, but
symmetry breaking of such solitons is generically forbidden
\cite{Yang_Stud14}. However, for certain special forms of 1D \PT
potentials, symmetry breaking of 1D solitons has been reported very
recently \cite{Yang_breaking}.

In this article, we show that symmetry breaking of 2D solitons is
also possible in the model (\ref{Eq:NLS}) for a special class of
complex potentials
\[  \label{Vform}
V(x,y)=g^2(x)+\alpha g(x)+ig'(x)  + h(y),
\]
where $g(x)$ is a real even function, i.e.,
\[ g(-x)=g(x),   \nonumber
\]
$h(y)$ is an arbitrary real function, and $\alpha$ is a real
constant. This potential is separable in $(x,y)$, and its imaginary
part is $y$-independent. In addition, this potential is
\PPT-symmetric, i.e.,
\[  \label{e:VPPT}
V^*(x, y)=V(-x, y),
\]
where the asterisk represents complex conjugation. Due to
separability of this potential, it is easy to see that its linear
spectrum can be all-real \cite{Yang_PPT}. Note that a potential of
the form (\ref{Vform}) but with $x$ and $y$ switched is equivalent
to (\ref{Vform}) and thus does not deserve separate consideration.

The $x$-component of the separable potential (\ref{Vform}) is the
same as the 1D complex potential for symmetry breaking as reported
in \cite{Yang_breaking}, but the $y$-component of this separable
potential is real and quite different. Should this $y$-component be
complex and also take the form of its $x$-component, we have found
that symmetry breaking would no longer occur. This indicates that
symmetry breaking in the special 2D potential (\ref{Vform}) is by no
means obvious and cannot be anticipated from the 1D potential for
symmetry breaking in \cite{Yang_breaking}.

Below we use two explicit examples of the potential (\ref{Vform}) to
demonstrate symmetry breaking of 2D solitons and reveal their unique
linear-stability properties.

\begin{figure}[!htbp]
\centering
\includegraphics[width=0.5\textwidth]{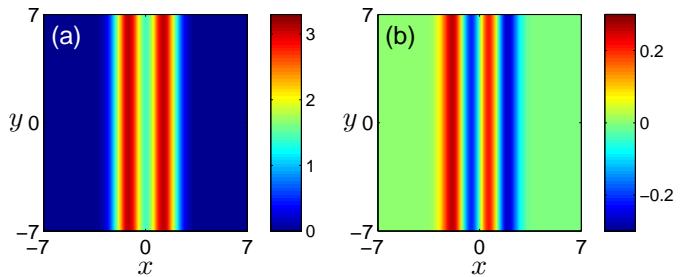}
\caption{A stripe complex potential (\ref{Vform}) with (\ref{e:gx})-(\ref{e:alphah}) in Example 1. (a) Re($V$); (b) Im($V$).} \label{f:fig1}
\end{figure}

\vspace{0.3cm} \textbf{Example 1 }\;  In our first example, we take
the potential (\ref{Vform}) with
\[  \label{e:gx}
g(x)=0.3\left[e^{-(x+1.2)^2}+e^{-(x-1.2)^2}\right],
\]
\[  \label{e:alphah}
\alpha=10, \quad h(y)=0.
\]
This is a $y$-independent stripe potential which is illustrated in
Fig.~\ref{f:fig1}. The spectrum of this potential is all-real, and
all eigenvalues lie in the continuous spectrum of $(-\infty,
\hspace{0.02cm} 2.0569]$.

Solitons in Eq. (\ref{e:soliton}) under this potential will be
computed by the Newton-conjugate-gradient method. This method
features high accuracy as well as fast speed. The application of
this method for solitons in conservative systems has been described
in \cite{Yang_2009,Yang_book}. In those cases, the linear
Newton-correction equation was self-adjoint and thus could be solved
directly by preconditioned conjugate gradient iterations. However,
the present equation (\ref{e:soliton}) is dissipative, hence the
resulting Newton-correction equation is non-self-adjoint. In this
case, direct conjugate gradient iterations on this equation would
fail, and it is necessary to turn this equation into a normal
equation and then solve it by preconditioned conjugate gradient
iterations. In the appendix, this Newton-conjugate-gradient method
for Eq. (\ref{e:soliton}) is explained in more detail. In addition,
a simple Matlab code is displayed.

\begin{figure}[!htbp]
\centering
\includegraphics[width=0.465\textwidth]{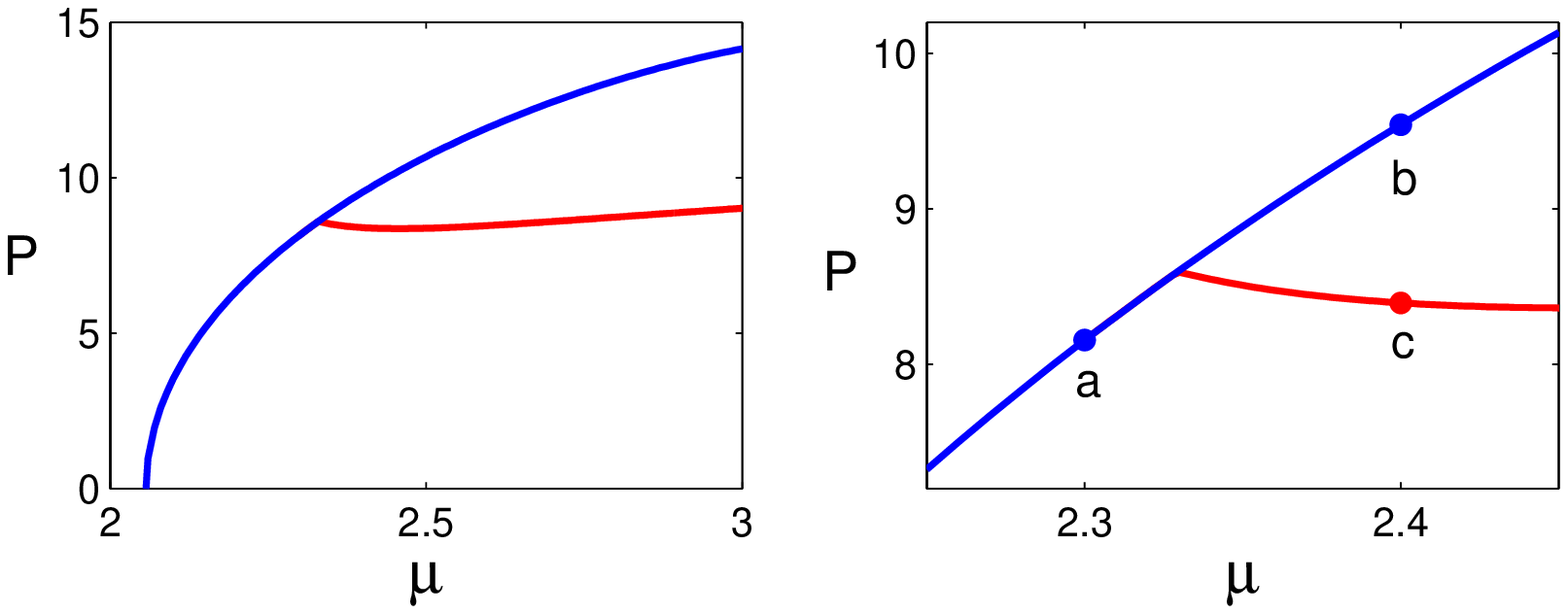}

\includegraphics[width=0.5\textwidth]{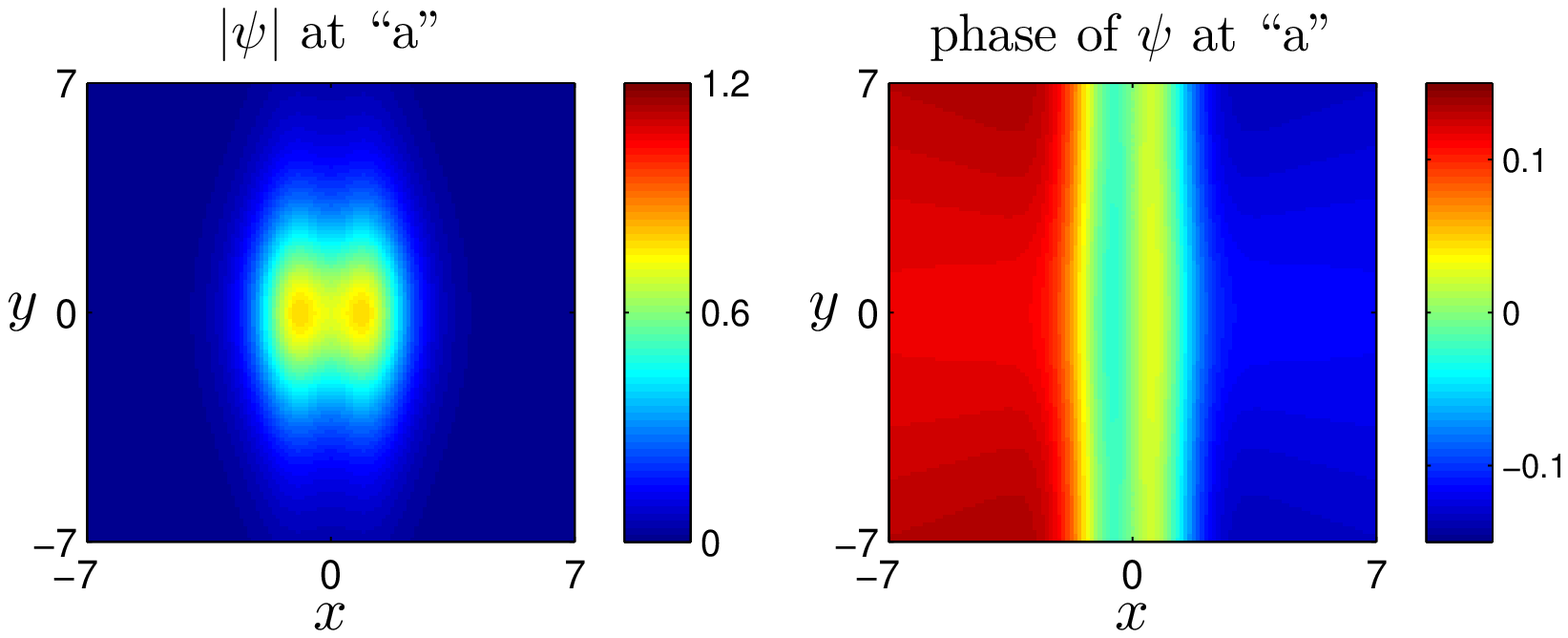}

\vspace{0.2cm}
\includegraphics[width=0.5\textwidth]{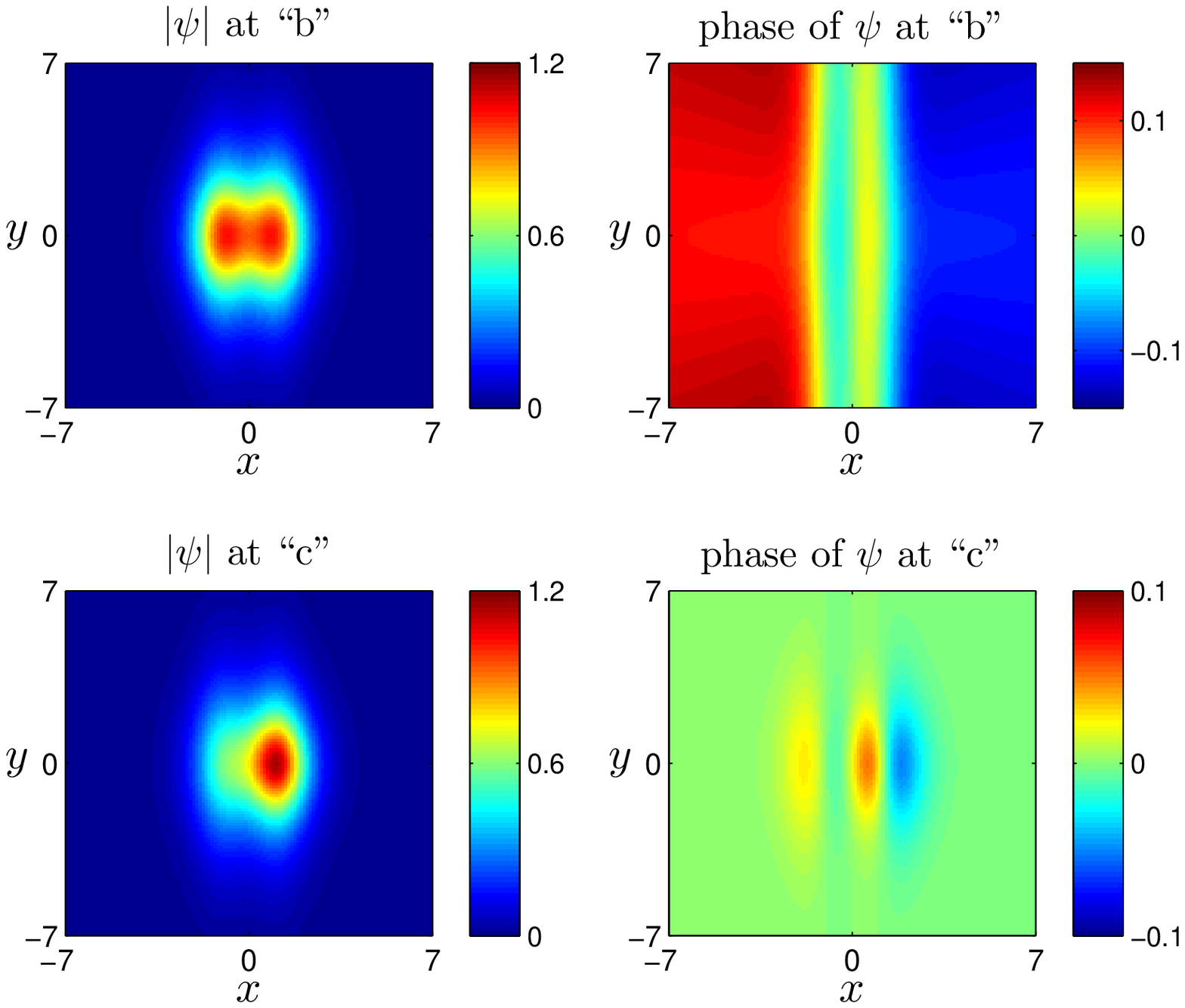}
\caption{Symmetry breaking of solitons in Example 1. First row:
power curves of symmetric (blue) and asymmetric (red) solitons; the
right panel is an amplification of the left panel around the
bifurcation point. Second to fourth rows: soliton profiles at points
`a,b,c' of the power curve; left panels: amplitude fields; right
panels: phase fields. } \label{f:fig2}
\end{figure}

Using this Newton-conjugate-gradient method, we find that from the
edge of the continuous spectrum $\mu_0=2.0569$, a continuous family
of solitons $\psi_s(x,y;\mu)$, localized in both $x$ and $y$
directions, bifurcate out. The power curve of this soliton family is
displayed in Fig.~\ref{f:fig2} (blue curve in the first row). Here
the power is defined as
\[
P(\mu)=\int_{-\infty}^\infty \int_{-\infty}^\infty |\psi(x,y;\mu)|^2 dxdy.  \nonumber
\]
At two points `a,b' of this power curve, soliton profiles are shown
in Fig.~\ref{f:fig2} (the second and third rows). These solitons
respect the same \PPT symmetry of the potential, i.e.,
\[ \label{psisym}
\psi_s^*(x,y)=\psi_s(-x,y).
\]
The existence of this soliton family respecting the same symmetry of
the potential is anticipated.

What is surprising is that, when the power of this soliton family
reaches a critical value $P_c\approx 8.60$, two branches of
asymmetric solitons bifurcate out through a pitchfork bifurcation.
These asymmetric solitons do not respect the \PPT symmetry
(\ref{psisym}). At the same $\mu$ value, they have identical powers
and are related to each other through a spatial reflection
\[  \label{mirrorsym}
\psi_a^{(1)*}(x, y)=\psi_a^{(2)}(-x, y).
\]
The power curve of these two branches of asymmetric solitons is
plotted in Fig.~\ref{f:fig2} (red curve in the first row). Notice
that unlike the symmetric (base) branch, the power slope of these
asymmetric branches is negative at the bifurcation point. At point
`c' of the asymmetric branches, the profile for one of the two
asymmetric solitons is displayed in Fig.~\ref{f:fig2} (the bottom
row). Asymmetry in its profile can clearly be seen. These solitons
have lost the \PPT symmetry of the underlying potential, thus
symmetry breaking has occurred.

Next we analyze linear stability of these symmetric and asymmetric
solitons. To determine linear stability, we perturb these solitons
as
\begin{equation}
\Psi(x,y,z) = e^{i\mu z} \left[ \psi(x,y) + \tilde{u}(x,y) \hspace{0.05cm}
e^{ \lambda z} + \tilde{w}^*(x,y) \hspace{0.05cm} e^{ \lambda^* z}
\right],  \nonumber
\end{equation}
where $|\tilde{u}|,  |\tilde{w}|\ll |\psi|$. After substitution into
equation (\ref{Eq:NLS}) and linearizing, we arrive at the eigenvalue
problem
\begin{equation}
{\cal L} \left(\begin{array}{c} \tilde{u}\\ \tilde{w}
\end{array} \right) = \lambda   \left(\begin{array}{c} \tilde{u}\\ \tilde{w}
\end{array}\right),  \label{EigenProblem}
\end{equation}
where
\[
{\cal L}= i\left( \begin{array}{c c} {\cal L}_{11} &  {\cal L}_{12} \\
{\cal L}_{21} & {\cal L}_{22}\end{array} \right),  \nonumber
\]
and
\begin{align*}
{\cal L}_{11} &= \nabla^2 + V -\mu  + 2 \sigma |\psi|^2,  \\
{\cal L}_{12} &= \sigma \psi^2, \\
{\cal L}_{21} &= -\sigma \left(\psi^2\right)^*,  \\
{\cal L}_{22} &= -\left(\nabla^2 + V -\mu  + 2 \sigma |\psi|^2\right)^*.
\end{align*}
If eigenvalues with positive real parts exist, the soliton is
linearly unstable; otherwise it is linearly stable.

Linear-stability eigenvalues exhibit important differences for
symmetric and asymmetric solitons. For symmetric solitons
$\psi_s(x,y)$, it is easy to show from soliton symmetry
(\ref{psisym}) and potential symmetry (\ref{e:VPPT}) that if
$$\lambda, \, \tilde{u}(x,y), \, \tilde{w}(x,y)$$
is an eigenmode, then so is
$$\lambda^*, \, \tilde{w}^*(x,y),  \, \tilde{u}^*(x,y),$$
$$-\lambda, \,  \tilde{w}(-x,y), \,  \tilde{u}(-x,y),$$
and
$$-\lambda^*, \,  \tilde{u}^*(-x,y), \, \tilde{w}^*(-x,y).$$
Thus for symmetric solitons, real and imaginary eigenvalues appear
as pairs $(\lambda, -\lambda)$, and complex eigenvalues appear as
quartets $\{\lambda, \lambda^*, -\lambda, -\lambda^*\}$.

For asymmetric solitons, however, the situation is different. While
it is still true that if $\lambda$ is an eigenvalue, so is
$\lambda^*$, but due to the lack of soliton symmetry (\ref{psisym}),
$-\lambda$ and $-\lambda^*$ are no longer eigenvalues. In other
words, for asymmetric solitons, complex eigenvalues appear as
conjugate pairs $(\lambda, \lambda^*)$, not as quartets; and real
eigenvalues appear as single eigenvalues, not as $(\lambda,
-\lambda)$ pairs. These differences on eigenvalue symmetry between
symmetric and asymmetric solitons will have important implications,
as we will see later in this section.

For the two branches of asymmetric solitons, their linear-stability
eigenvalues are related. Indeed, from the mirror symmetry
(\ref{mirrorsym}) between these two bifurcated soliton branches, it
is easy to see that if $\lambda$ is an eigenvalue of the soliton
$\psi_a^{(1)}(x, y; \mu)$, then $-\lambda^*$ will be an eigenvalue
of the companion soliton $\psi_a^{(2)}(x, y; \mu)$. In other words,
linear-stability spectrum of the soliton $\psi_a^{(1)}(x, y; \mu)$
is a mirror reflection of that spectrum of the companion soliton
$\psi_a^{(2)}(x, y; \mu)$ around the imaginary axis.

\begin{figure}[!htbp]
\centering
\includegraphics[width=0.5\textwidth]{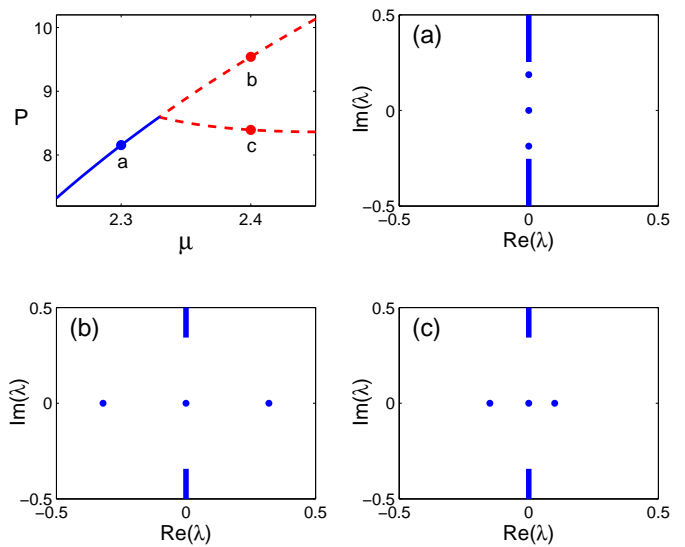}
\caption{Linear-stability behaviors of solitons near the symmetry-breaking point in Example 1.
Upper left panel: the power curve with stability marked (solid blue for stable and dashed red for unstable). Panels (a,b,c): linear-stability spectra for
the solitons at points `a,b,c' of the power curve. } \label{f:fig3}
\end{figure}

The eigenvalue problem (\ref{EigenProblem}) can be computed by the
Fourier collocation method (for the full spectrum) or the
Newton-conjugate-gradient method (for individual discrete
eigenvalues) \cite{Yang_book}. We find that near the
symmetry-breaking bifurcation point $\mu_c\approx 2.33$, symmetric
solitons are stable before the bifurcation point ($\mu<\mu_c$) and
unstable after it ($\mu>\mu_c$), and both branches of asymmetric
solitons are unstable. This stability behavior is marked on the
power curve in Fig. \ref{f:fig3} (upper left panel). To shed light
on the origins of these stabilities and instabilities,
linear-stability spectra at three points `a,b,c' of this power
curve, for the three solitons displayed in Fig.~\ref{f:fig2}, are
displayed in panels (a,b,c) of Fig.~\ref{f:fig3} respectively. We
see from panel (a) that before the bifurcation, the symmetric
soliton has a pair of discrete eigenvalues on the imaginary axis. At
the bifurcation point, this pair of imaginary eigenvalues coalesce
at the origin. After bifurcation, these coalesced eigenvalues split
along the real axis in opposite directions for both symmetric and
asymmetric solitons. Along the symmetric branch, the two split
eigenvalues form a $(\lambda, -\lambda)$ pair [see panel (b)]. But
along the asymmetric branches, the two split eigenvalues do not form
a $(\lambda, -\lambda)$ pair since they have different magnitudes
[see panel (c)]. These spectra show that the instability of
symmetric and asymmetric solitons after bifurcation is due to the
zero-eigenvalue splitting along the real axis at $\mu=\mu_c$, and
this instability is exponential (caused by real eigenvalues).

It is interesting to observe that the power-curve structure and the
associated stability behaviors in Fig. \ref{f:fig3} (upper left
panel) resemble that in the conservative generalized nonlinear
Schr\"odinger equations with real potentials (see Fig. 2c in Ref.
\cite{YangPhyd}). In that conservative case, it was shown that if
the power slopes of the symmetric and asymmetric solitons at the
bifurcation point have opposite signs, then both solitons will share
the same stability or instability \cite{YangPhyd}. Fig. \ref{f:fig3}
of the present article suggests that such a statement might hold for
complex potentials as well. But whether it holds for other complex
potentials merits further investigation.

\begin{figure}[!htbp]
\centering
\includegraphics[width=0.5\textwidth]{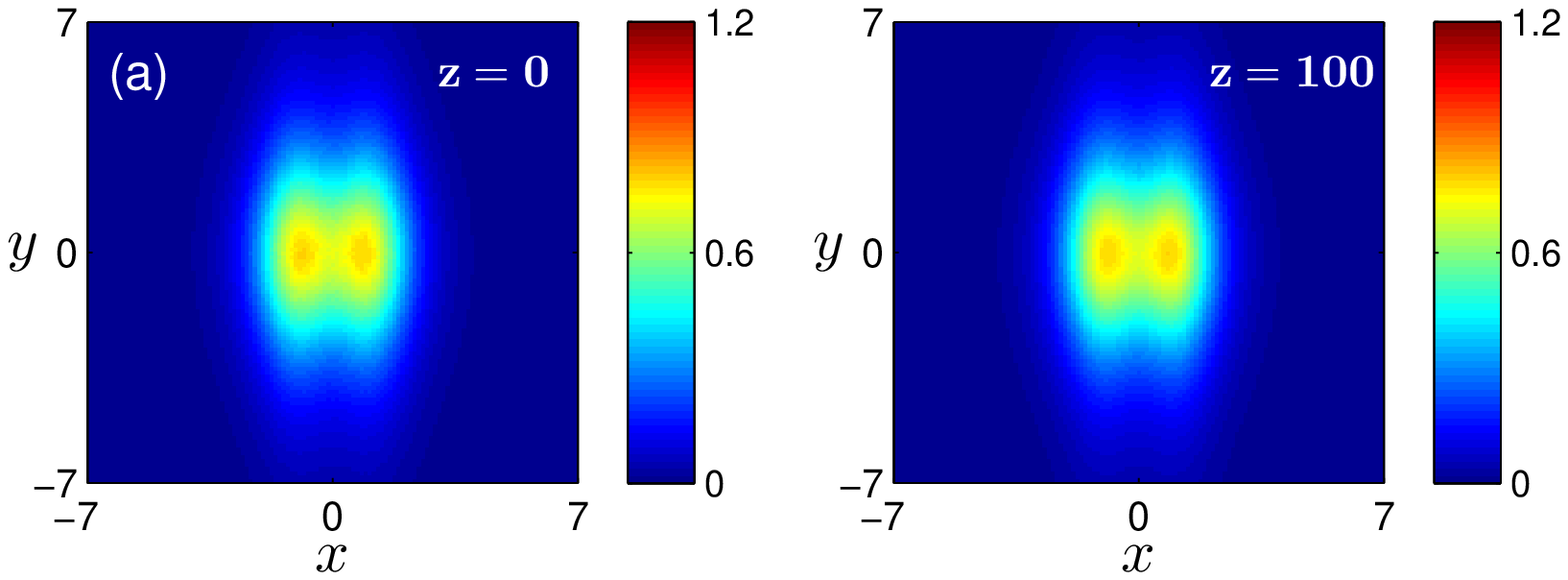}

\vspace{0.2cm}
\includegraphics[width=0.5\textwidth]{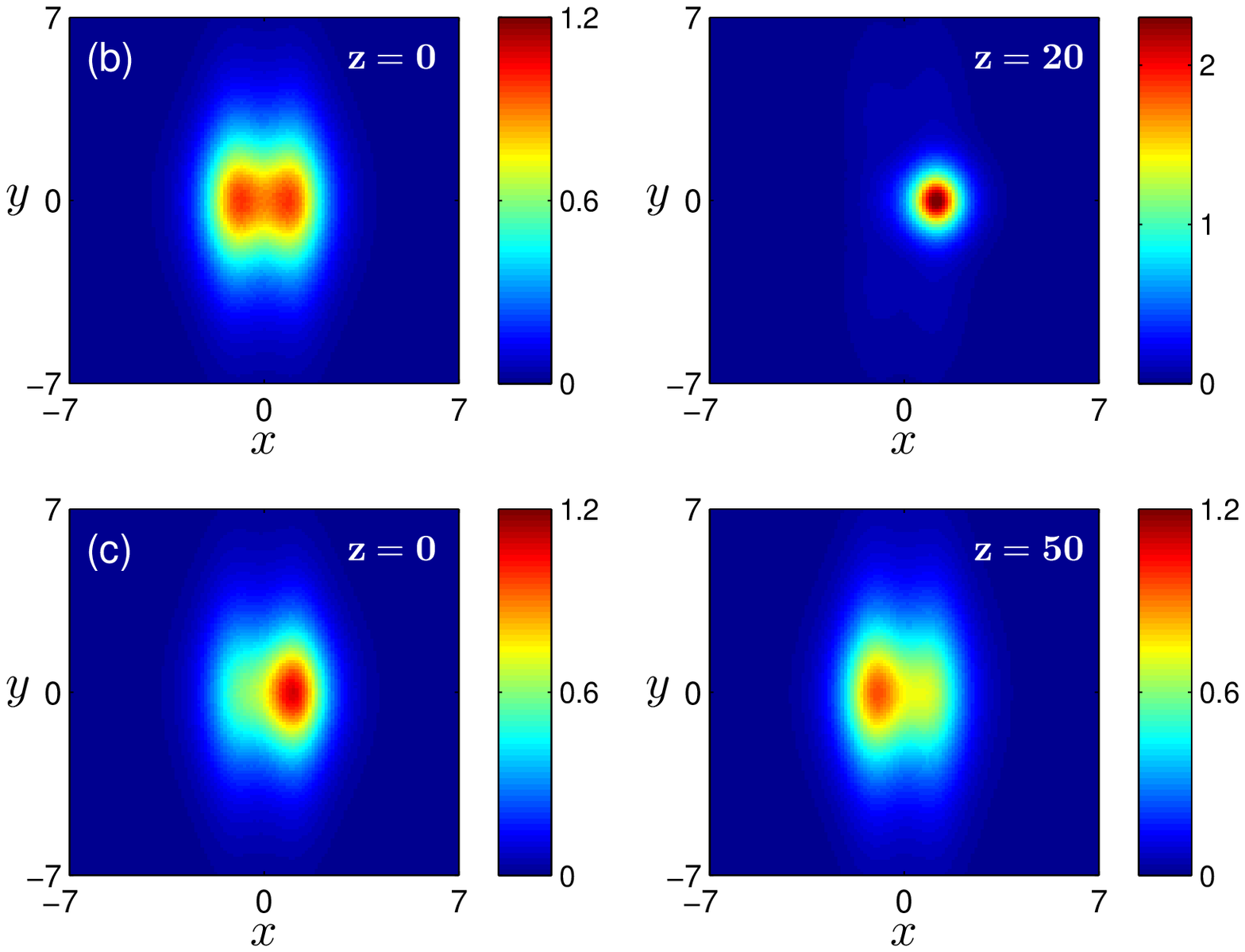}
\caption{Nonlinear evolutions of the three solitons in Fig.
\ref{f:fig2} under 1\% random-noise perturbations (locations of
these solitons on the power curve are marked in both Figs.
\ref{f:fig2} and \ref{f:fig3}). } \label{f:fig4}
\end{figure}

The linear-stability results of Fig.~\ref{f:fig3} are corroborated
by nonlinear evolution simulations of those solitons under
random-noise perturbations. To demonstrate, we perturb the three
solitons of Fig.~\ref{f:fig2} by 1\% random-noise perturbations, and
their nonlinear evolutions are displayed in Fig.~\ref{f:fig4}. As
can be seen, the perturbed symmetric soliton before bifurcation
shows little change even after $z=100$ units of propagation,
confirming that it is linearly stable (see top row of
Fig.~\ref{f:fig4}). The perturbed symmetric soliton after
bifurcation, on the other hand, clearly breaks up and evolves into a
highly asymmetric profile after 20 units of propagation, confirming
that it is linearly unstable (see middle row of Fig.~\ref{f:fig4}).
The perturbed asymmetric soliton, whose initial intensity hump is
located at the right side, also breaks up and evolves into a profile
whose intensity hump moves to the left side after 50 units of
propagation, confirming that it is linearly unstable as well (see
bottom row of Fig.~\ref{f:fig4}).

\begin{figure}[!htbp]
\centering
\includegraphics[width=0.5\textwidth]{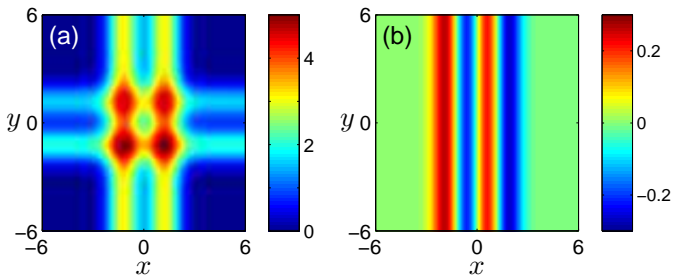}
\caption{The \PPT-symmetric complex potential (\ref{Vform}) in Example 2. (a) Re($V$); (b) Im($V$). } \label{f:fig5}
\end{figure}

\vspace{0.3cm} \textbf{Example 2 }\;  In our second example, we take
the potential (\ref{Vform}) with
\[
g(x)=0.3\left[e^{-(x+1.2)^2}+e^{-(x-1.2)^2}\right],  \quad \alpha=10,  \nonumber
\]
and
\[
h(y)=2\left[e^{-(y+1.2)^2}+0.8\hspace{0.02cm}  e^{-(y-1.2)^2}\right].   \nonumber
\]
This potential is illustrated in Fig.~\ref{f:fig5}. Its real part is
no longer a stripe potential, neither is it symmetric in $y$. The
spectrum of this potential is all-real, and it consists of three
discrete eigenvalues of $\{2.5643, 2.5689, 3.2028\}$ and the
continuous spectrum of $(-\infty, 2.0569]$.

\begin{figure}[!htbp]
\centering
\includegraphics[width=0.485\textwidth]{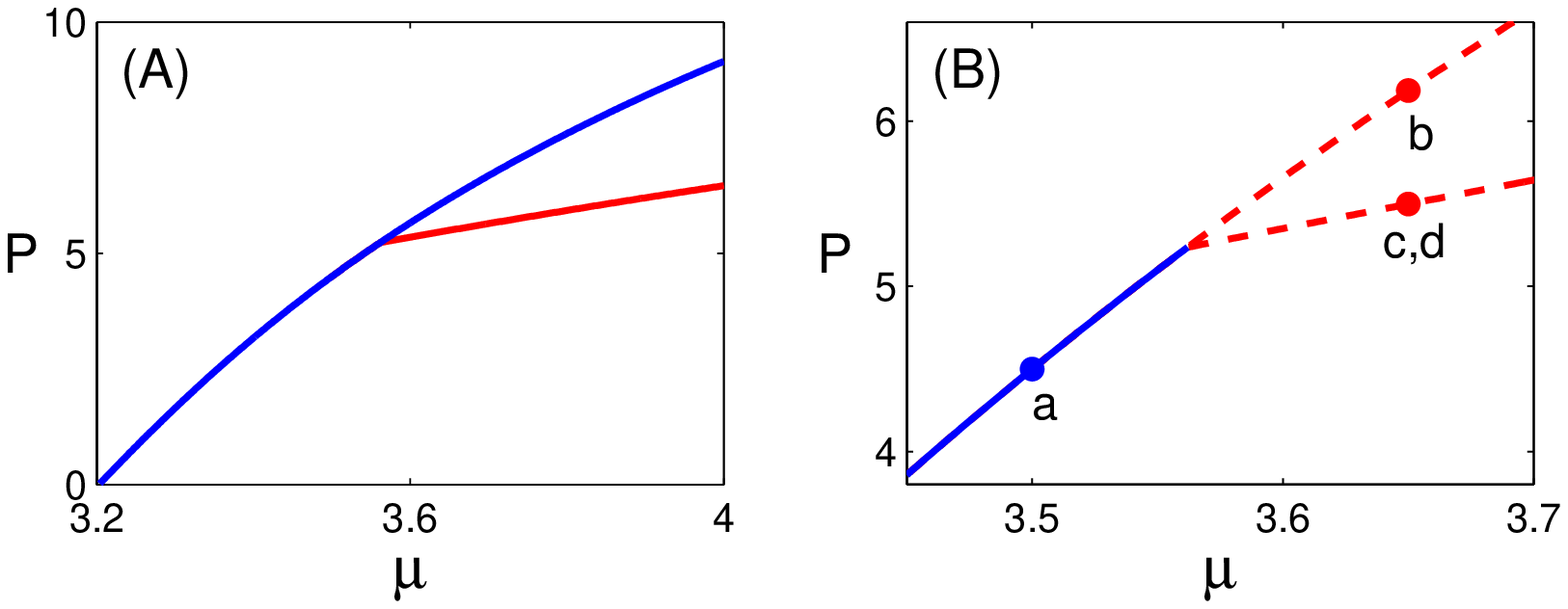}

\vspace{0.2cm}
\includegraphics[width=0.5\textwidth]{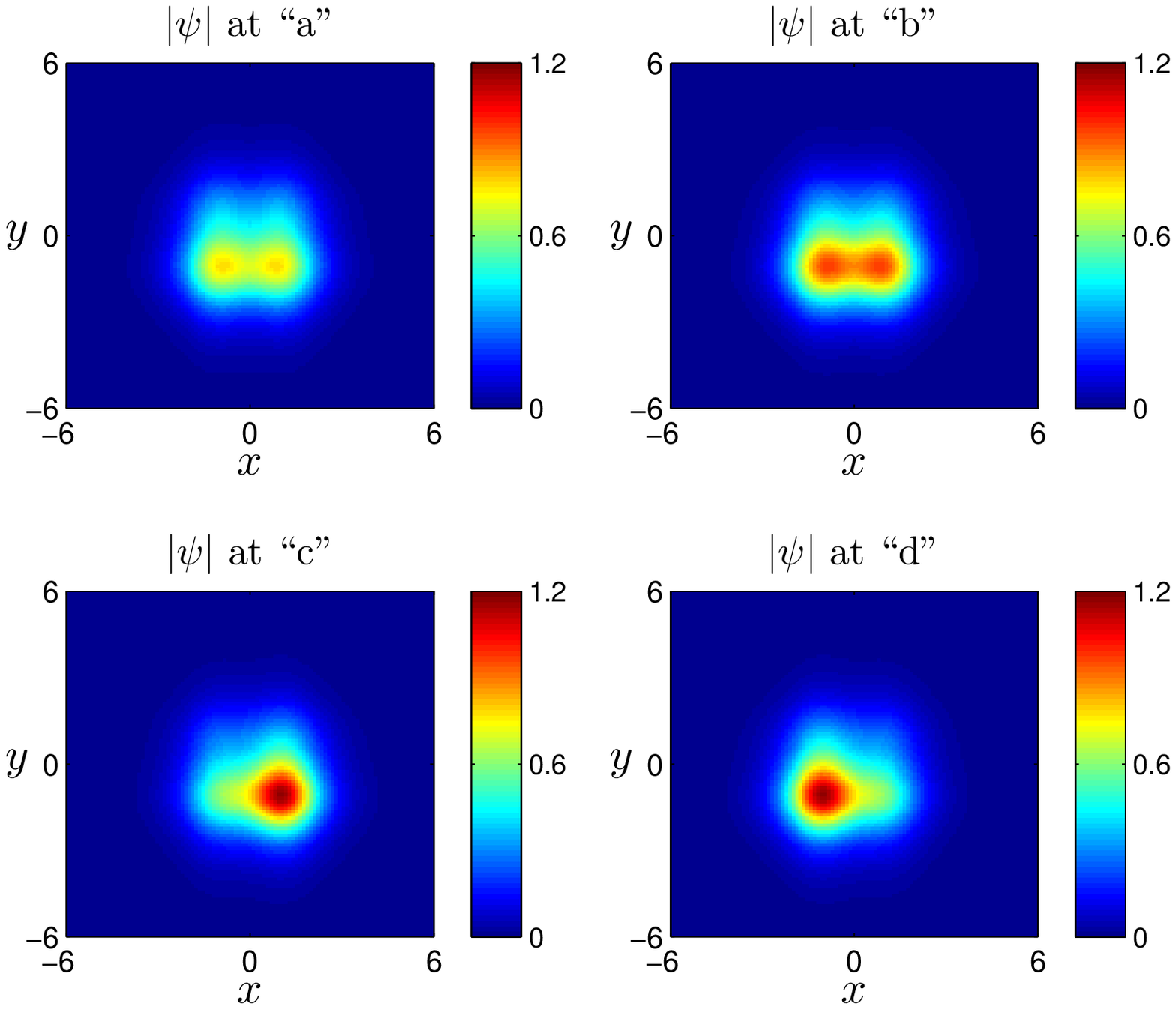}
\caption{Symmetry breaking of solitons in Example 2. (A) power
curves of \PPT-symmetric (blue) and asymmetric (red) solitons; (B)
enlargement of the left power curves near the bifurcation point
(solid blue indicates linearly-stable branch, and dashed red
indicates linearly-unstable branches). Middle and bottom rows:
profiles of soliton amplitudes at points `a,b,c,d' of the power
curve.  } \label{f:fig6}
\end{figure}

From the largest discrete eigenvalue of $\mu_0=3.2028$, a continuous
family of \PPT-symmetric solitons bifurcates out. The power curve of
this soliton family is plotted in Fig.~\ref{f:fig6}(A) (blue curve).
When the power of these solitons reaches a threshold of $P_c\approx
5.24$ (at $\mu_c\approx 3.56$), two branches of asymmetric solitons
bifurcate out, whose power curves are also displayed in
Fig.~\ref{f:fig6}(A) (red curve). As before, these two asymmetric
solitons are related to each other by
\[  \label{sym20}
\psi_a^{(1)*}(x, y)=\psi_a^{(2)}(-x, y),
\]
thus they have identical powers. Enlargement of this power curve
near the bifurcation point is shown in Fig.~\ref{f:fig6}(B). At
points `a,b,c,d' of this amplified power diagram, the solitons'
amplitude profiles are plotted in Fig.~\ref{f:fig6} (middle and
bottom rows). Here points `c,d' are the same power points but on
different asymmetric-soliton branches. We can see that solitons at
points `a,b' of the base branch are \PPT-symmetric, with `a' before
bifurcation and `b' after it. The solitons at point `c,d' of the
bifurcated branches, however, are asymmetric, with the energy
concentrated on the right and left side of the $x$-axis
respectively. In this example, power slopes of the base and
bifurcated soliton branches have the same sign at the bifurcation
point, which is different from Example~1.

\begin{figure}[!htbp]
\centering
\includegraphics[width=0.5\textwidth]{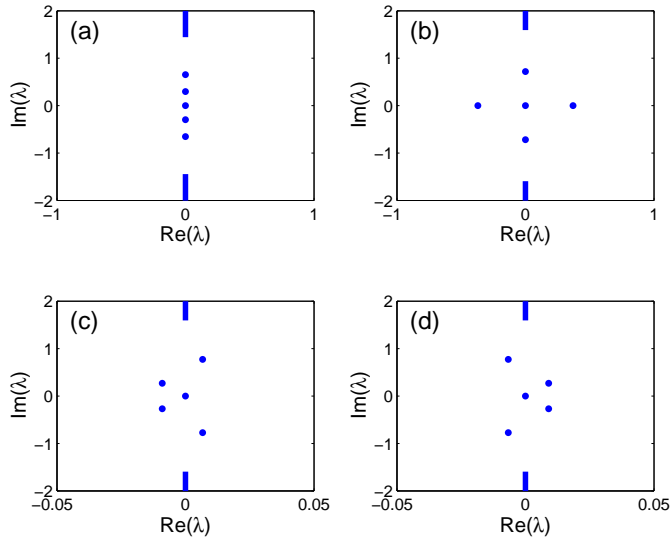}
\caption{(a-d): Linear-stability spectra for solitons at points `a-d' of the power curve in Fig.~\ref{f:fig6}(B).} \label{f:fig7}
\end{figure}

Now we discuss linear-stability behaviors of solitons in Example 2.
For the base branch of \PPT-symmetric solitons, they are linearly
stable before the bifurcation point and linearly unstable after it,
which is similar to Example 1 and is not surprising. To illustrate,
linear-stability spectra for the two \PPT-symmetric solitons at
points `a,b' of the power curve in Fig.~\ref{f:fig6}(B) are plotted
in Fig.~\ref{f:fig7}(a,b) respectively. At point `a' (before
bifurcation), all eigenvalues are imaginary, indicating linear
stability. At point `b' (past bifurcation), a pair of real
eigenvalues $\pm 0.3704$ appear, which makes this \PPT-symmetric
soliton linearly unstable. What happens is that when the power of
the base branch crosses the bifurcation point, a pair of imaginary
eigenvalues collide at the origin and then bifurcate out of the
origin along the real axis, creating a $\pm \lambda$  pair of real
eigenvalues and hence instability.

The most interesting new phenomena in Example 2 are linear-stability
behaviors of asymmetric solitons. We find that both branches of
asymmetric solitons are linearly unstable, but origins of their
instabilities are different. To demonstrate, linear-stability
spectra for the two asymmetric solitons at points `c,d' of
Fig.~\ref{f:fig6}(B) are plotted in Fig.~\ref{f:fig7}(c,d). These
two spectra are related to each other by mirror reflection around
the imaginary axis, as we have pointed out earlier in the text. In
addition, eigenvalues of these asymmetric solitons must appear in
conjugate pairs $(\lambda, \lambda^*)$, but no other eigenvalue
symmetry exists.

The first phenomenon we notice in these spectra is that, both
asymmetric solitons are linearly unstable due to oscillatory
instabilities caused by complex eigenvalues. The second phenomenon
is that, even though these spectra contain complex eigenvalues,
these eigenvalues do \emph{not} appear in quartets $\{\lambda,
\lambda^*, -\lambda, -\lambda^*\}$. This contrasts asymmetric
solitons in real (conservative) potentials, where complex
eigenvalues must appear in quartets.

The third and probably most noteworthy phenomenon in these spectra
is that, unstable eigenvalues in these two asymmetric solitons have
different origins. Indeed, before the bifurcation, \PPT-symmetric
solitons on the base branch have two pairs of simple discrete
imaginary eigenvalues [see Fig.~\ref{f:fig7}(a)]. At the bifurcation
point, the smaller pair of simple imaginary eigenvalues coalesce at
the origin, while the larger pair remain on the imaginary axis. When
asymmetric solitons bifurcate out from the base branch, for the one
with energy concentrated on the right side (see Fig.~\ref{f:fig6},
at point `c'), the pair of simple eigenvalues on the imaginary axis
move directly to the right half plane, creating oscillatory
instability [see Fig.~\ref{f:fig7}(c)]. The coalesced zero
eigenvalues at the origin, on the other hand, move leftward into the
complex plane, creating a conjugate pair of stable complex
eigenvalues [see Fig.~\ref{f:fig7}(c)]. For the asymmetric soliton
with energy concentrated on the left side, the situation is just the
opposite [see Fig.~\ref{f:fig7}(d)]. Thus the origin of instability
for one branch of asymmetric solitons is due to a pair of simple
imaginary eigenvalues moving directly off the imaginary axis, while
the origin for the other branch of asymmetric solitons is due to the
zero eigenvalue moving to the complex plane.

The above phenomenon of zero and simple imaginary eigenvalues moving
directly into the complex plane and creating oscillatory instability
in solitons is very novel, since it contrasts conservative systems
with real potentials. In real potentials, linear-stability complex
eigenvalues of solitons appear as quartets $\{\lambda, \lambda^*,
-\lambda, -\lambda^*\}$. Partly because of it, bifurcation of
complex eigenvalues off the imaginary axis typically occurs through
collision of imaginary eigenvalues of opposite Krein signatures (a
bifurcation referred to as Hamiltonian-Hopf bifurcation in the
literature \cite{Peli_Hopf}). In addition, complex eigenvalues (not
on the real and imaginary axes) cannot bifurcate from the origin
when two simple eigenvalues collide there. But in complex
potentials, the situation can be very different as is explained
above.

The fourth phenomenon in the spectra of Fig.~\ref{f:fig7} is that,
the maximal growth rates of perturbations in these two asymmetric
solitons are different. Indeed the unstable eigenvalues in
Fig.~\ref{f:fig7}(c) are $0.0067\pm 0.7721i$, giving a growth rate
of $0.0067$; while the unstable eigenvalues in Fig.~\ref{f:fig7}(d)
are $0.0090 \pm 0.2692i$, giving a larger growth rate of 0.0090. The
fifth phenomenon is that these oscillatory instabilities in
asymmetric solitons are rather weak due to these small growth rates.
This means that these oscillatory instabilities will take long
distances to develop.

Of the five phenomena mentioned above, the third and fourth ones are
the most fundamental, and they are rarely seen (if ever) for
asymmetric solitons arising from symmetry-breaking bifurcations.

\begin{figure}[!htbp]
\centering
\includegraphics[width=0.5\textwidth]{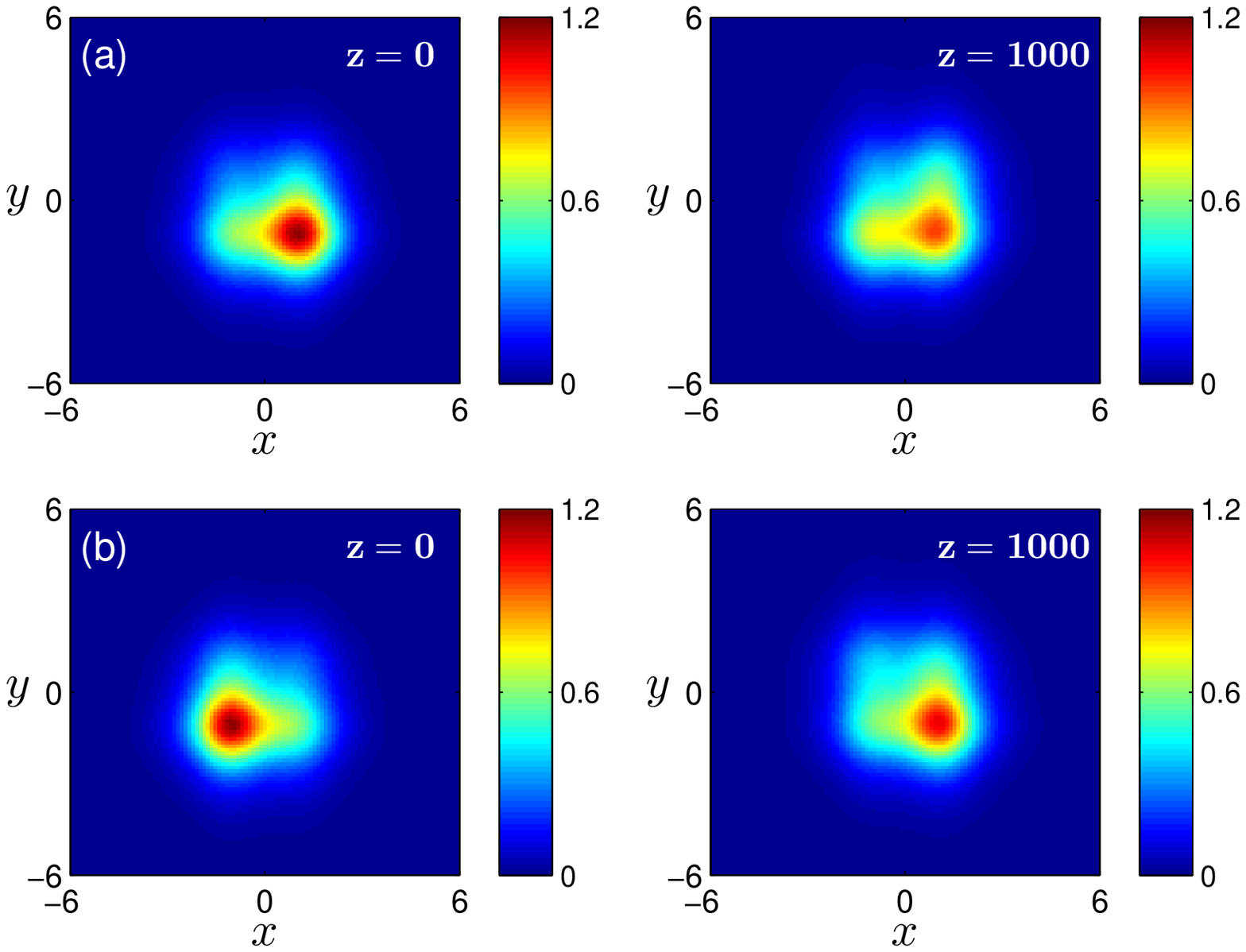}

\vspace{0.2cm}
\includegraphics[width=0.5\textwidth]{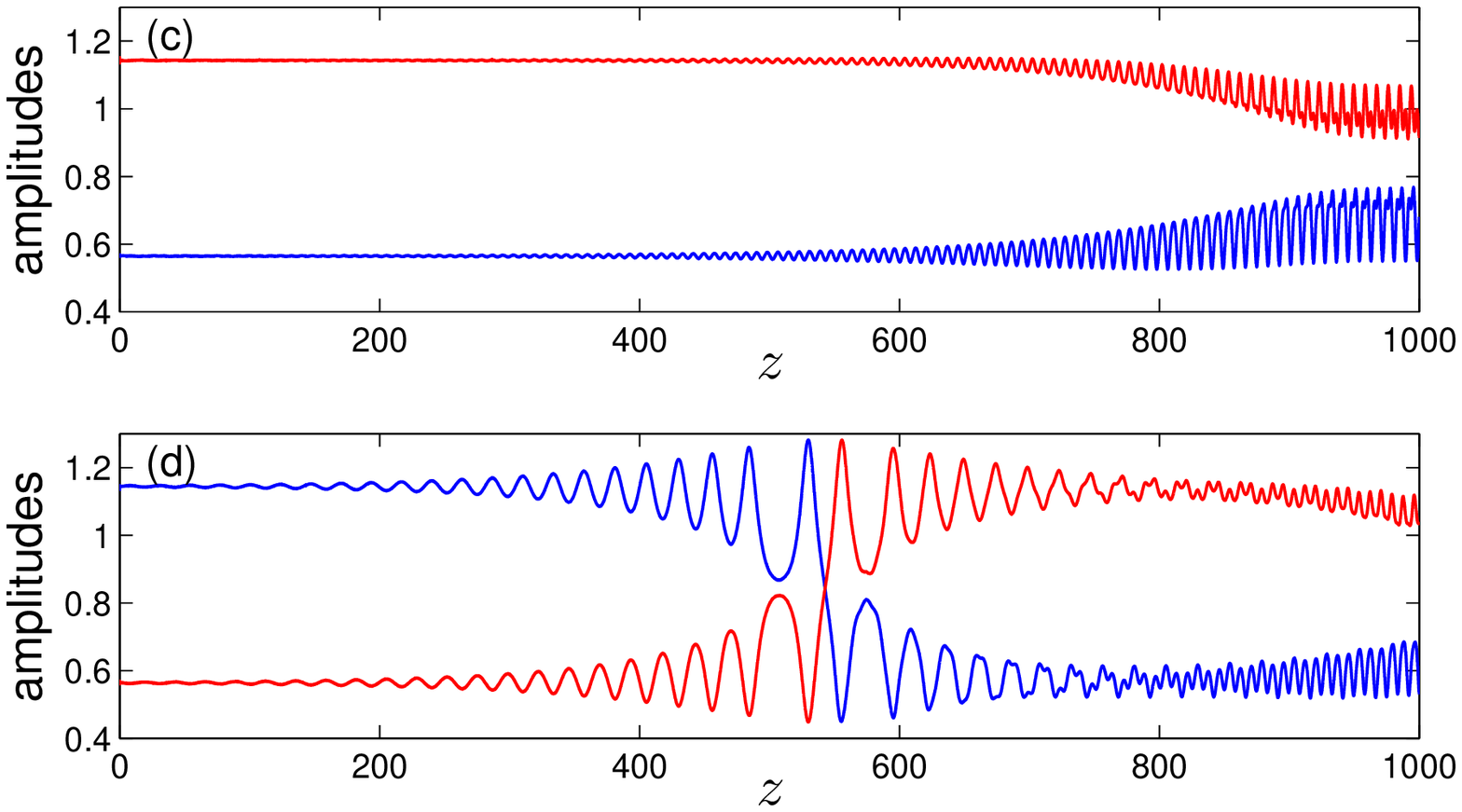}
\caption{Non-reciprocal evolutions of two reciprocal asymmetric
solitons in Fig.~\ref{f:fig6} under 1\% random-noise perturbations
in Example~2. First and second rows: initially perturbed asymmetric
solitons and their evolved solutions at $z=1000$. (c,d) Evolutions
of solution amplitudes $|\Psi|$ versus $z$ at two spatial positions
$(x,y)=(-1.2, -1.2)$ (blue) and $(1.2, -1.2)$ (red) for the two
asymmetric solitons of Fig.~\ref{f:fig6} under perturbations. }
\label{f:fig8}
\end{figure}

Since the two branches of asymmetric solitons have different origins
of instability and different growth rates, small perturbations in
these solitons will grow differently, leading to non-reciprocal
developments of instability. To demonstrate, evolutions of the two
asymmetric solitons in Fig.~\ref{f:fig6} under 1\% random-noise
perturbations are displayed in Fig.~\ref{f:fig8}. We see that even
though these two asymmetric solitons are related to each other by a
mirror reflection (\ref{sym20}) and are reciprocal, their evolutions
under weak perturbations are not reciprocal. Indeed, after 1000
distance units of propagation, they reach similar asymmetric states.
This non-reciprocal evolution is most visible in
Fig.~\ref{f:fig8}(c,d), where amplitude evolutions at spatial
positions $(x,y)=(-1.2, -1.2)$ and $(1.2, -1.2)$ for the two
perturbed asymmetric solitons are plotted respectively. These
amplitude evolutions vividly confirm that (a) the two asymmetric
solitons are linearly unstable; (b) their instabilities are caused
by different unstable modes with different growth rates; and (c) the
nonlinear evolutions are non-reciprocal even though the asymmetric
solitons are.

In Example 2, when asymmetric solitons bifurcate out, the coalesced
zero eigenvalue and the pair of imaginary eigenvalues move in
opposite directions in the complex plane, causing instability to
both asymmetric solitons [see Fig.~\ref{f:fig7}(c,d)]. For other
potentials and/or nonlinearities, if those eigenvalues bifurcate in
the same direction, then one asymmetric soliton would be linearly
stable and the other unstable. Such a scenario would be very
remarkable. Whether such scenarios exist or not is an open question.

In the above two examples, symmetry breaking was observed for
complex potentials of the form (\ref{Vform}). We have also tried a
related class of complex potentials
\[  \label{Vform2}
V(x,y)=g^2(x)+\alpha g(x)+ig'(x)  + h^2(y)+\beta \hspace{0.03cm} h(y)+ ih'(y),
\]
where $g(x), h(y)$ are real even functions, and $\alpha, \beta$ are
real constants. This potential is \PT-symmetric, i.e.,
$V^*(x,y)=V(-x,-y)$, and it admits \PT-symmetric solitons. But we
did not find symmetry breaking here, i.e., we did not find branches
of asymmetric solitons bifurcating from the branch of \PT-symmetric
solitons.

Why does symmetry breaking occur in potentials of the form
(\ref{Vform}) but not in some others such as (\ref{Vform2})? This
question is not clear yet. In fact, even for one-dimensional
symmetry-breaking bifurcations reported in \cite{Yang_breaking}, the
reason for that symmetry breaking was not entirely clear either. In
the 1D case, the forms of potentials for symmetry breaking in
\PT-symmetric potentials and for soliton families in asymmetric
potentials are the same \cite{Yang_breaking,Tsoy2014}. For those
potentials, there is a conserved quantity which, when combined with
a shooting argument, helps explain the existence of soliton families
in asymmetric complex potentials \cite{Konotop_OL2004}. That
conserved quantity may prove useful to explain symmetry breaking in
those 1D potentials as well.

For the present class of 2D potentials (\ref{Vform}), we have found
that Eq. (\ref{Eq:NLS}) also admits a conservation law
\[
Q_t+J_{x}+K_{y}=0,
\]
where
\[
Q=i\Psi(\Psi_x^*-i\hat{g}\Psi^*),   \nonumber
\]
\[
J=\Psi\Psi_{yy}^*+|\Psi_x+i\hat{g}\Psi|^2-i\Psi\Psi_t^*+\left(h-\frac{\alpha^2}{4}\right)|\Psi|^2+\frac{\sigma}{2}|\Psi|^4,  \nonumber
\]
\[
K=\Psi_y(\Psi_x^*-i\hat{g}\Psi^*)-\Psi(\Psi_x^*-i\hat{g}\Psi^*)_y,   \nonumber
\]
and
\[ \hat{g}(x)=g(x)+\frac{\alpha}{2}.  \nonumber
\]
For solitons (\ref{soliton}), substituting their functional form
into the above conservation law, a reduced conservation law for the
soliton function $\psi(x,y)$ can also be derived. For the other
class of potentials (\ref{Vform2}), however, we could not find such
a conservation law. This suggests that there is indeed a connection
between the existence of a conservation law and the presence of
symmetry breaking of solitons. But this connection in the 2D case
would be harder to establish since shooting-type arguments would
break down.

In 1D, symmetry breaking in symmetric potentials and existence of
soliton families in asymmetric potentials occur in complex
potentials of the same form \cite{Yang_breaking,Tsoy2014}. This
invites a natural question: for the class of 2D complex potentials
(\ref{Vform}) which admits symmetry breaking, if these potentials
are not \PPT-symmetric, i.e., if $g(x)$ is real but not even, can
they support continuous families of solitons? The answer is positive
as our preliminary numerics has shown.

\section{Summary and Discussion}
In this article, we reported symmetry breaking of solitons in the
nonlinear Schr\"odinger equation with a class of two-dimensional
\PPT-symmetric complex potentials (\ref{Vform}). At the bifurcation
point, two branches of asymmetric solitons bifurcate out from the
base branch of \PPT-symmetric solitons, and this bifurcation is
quite surprising. Stability of these solitons near the bifurcation
point were also studied. In the two examples we investigated, we
found that the base branch of symmetric solitons changes stability
at the bifurcation point, and the bifurcated asymmetric solitons are
unstable. For the asymmetric solitons, two novel stability
properties were further revealed. One is that at the bifurcation
point, the zero and simple imaginary linear-stability eigenvalues of
asymmetric solitons can move directly into the complex plane and
create oscillatory instability. The other is that the two bifurcated
asymmetric solitons, even though having identical powers and being
related to each other by spatial mirror reflection, can have
different origins of linear instability and thus exhibit
non-reciprocal nonlinear evolutions under random-noise
perturbations.

We should point out that the complex potentials (\ref{Vform})
possess a single (\PPT) symmetry, thus they must be in that special
form in order for symmetry breaking to occur. If a complex potential
exhibits more than one spatial symmetry, say double \PPT symmetries
\[
V^*(x, y)=V(-x, y), \quad V^*(x, y)=V(x, -y),   \nonumber
\]
or one \PT and one \PPT symmetry, say
\[  \label{symI}
V^*(x, y)=V(-x, -y), \quad V^*(x, y)=V(-x, y),  \nonumber
\]
then this potential can admit symmetry breaking without the need for
special functional forms (this prospect has been mentioned in
\cite{Yang_Stud14} and confirmed by our own numerics). When symmetry
breaking occurs in such double-symmetry potentials, the base branch
of solitons respect both symmetries of the potential, while the
bifurcated solitons lose one symmetry but retain the other. The
simple mathematical reason for symmetry breakings in double-symmetry
potentials is that the infinitely many analytical conditions for
symmetry breaking in \cite{Yang_Stud14} are all satisfied
automatically due to the remaining symmetry of the bifurcated
solitons. That situation is fundamentally different from symmetry
breakings in potentials of special forms such as (\ref{Vform}),
which admit a single spatial symmetry. The mathematical reason for
symmetry breaking in single-symmetry potentials of special
functional forms such as (\ref{Vform}) is still not clear.

\section*{Acknowledgment}
This work was supported in part by the Air Force Office of
Scientific Research (Grant USAF 9550-12-1-0244) and the National
Science Foundation (Grant DMS-1311730).

\section*{Appendix: A Numerical Method for Computing Solitons in Complex Potentials}
\renewcommand{\theequation}{A.\arabic{equation}}

In this appendix, we describe the Newton-conjugate-gradient method
for computing solitons in Eq. (\ref{e:soliton}) with a complex
potential.

The general idea of the Newton-conjugate-gradient method is that,
for a nonlinear real-valued vector equation
\begin{equation} \label{L0U_CG}
\textbf{L}_0 (\textbf{u})=0,
\end{equation}
its solution $\textbf{u}$ is obtained by Newton iterations
\begin{equation} \label{Newtonit}
\textbf{u}_{n+1}=\textbf{u}_{n}+\Delta\textbf{u}_{n},
\end{equation}
where the updated amount $\Delta\textbf{u}_{n}$ is computed from the
linear Newton-correction equation
\begin{equation} \label{L1Dun}
\textbf{L}_{1n} \Delta\textbf{u}_n = -\textbf{L}_0 (\textbf{u}_n)
\end{equation}
where $\textbf{L}_{1n}$ is the linearization operator
$\textbf{L}_{1}$ of Eq. (\ref{L0U_CG}) evaluated at the approximate
solution $\textbf{u}_n$. If $\textbf{L}_{1}$ is self-adjoint, then
Eq. (\ref{L1Dun}) can be solved directly by preconditioned
conjugate-gradient iterations \cite{Yang_2009,Yang_book,Golub}. But
if $\textbf{L}_{1}$ is non-self-adjoint, we first multiply it by the
adjoint operator of $\textbf{L}_{1}$ and turn it into a normal
equation
\begin{equation} \label{L1Dun2}
\textbf{L}_{1n}^A \textbf{L}_{1n} \Delta\textbf{u}_n = -\textbf{L}_{1n}^A \textbf{L}_0 (\textbf{u}_n),
\end{equation}
which is then solved by preconditioned conjugate gradient
iterations.

For Eq. (\ref{e:soliton}), we first split the complex function
$\psi$ and the complex potential $V$ into their real and imaginary
parts,
\[
\psi=\psi_1+i\psi_2, \quad  V=V_1+iV_2.  \nonumber
\]
Substituting these equations into (\ref{e:soliton}), we obtain two
real equations for $(\psi_1, \psi_2)$ as
\begin{eqnarray*}
\nabla^2 \psi_1+(V_1-\mu)\psi_1-V_2\psi_2+\sigma(\psi_1^2+\psi_2^2)\psi_1=0, \\
\nabla^2\psi_2+(V_1-\mu)\psi_2+V_2\psi_1+\sigma(\psi_1^2+\psi_2^2)\psi_2=0.
\end{eqnarray*}
These two real equations are the counterpart of Eq. (\ref{L0U_CG})
for the vector function $\u=[\psi_1, \psi_2]^T$, where the
superscript `$T$' represents transpose of a vector. The
linearization operator of the above nonlinear equations is
\[
\textbf{L}_1=\left[ \begin{array}{cc}  L_{11} & L_{12} \\
L_{21} & L_{22} \end{array}\right],   \nonumber
\]
where
\begin{eqnarray*}
L_{11} & = & \nabla^2+V_1-\mu+\sigma(3\psi_1^2+\psi_2^2), \\
L_{12} & = & 2\sigma\psi_1\psi_2-V_2, \\
L_{21} & = & 2\sigma\psi_1\psi_2+V_2, \\
L_{22} & = & \nabla^2+V_1-\mu+\sigma(3\psi_2^2+\psi_1^2).
\end{eqnarray*}
This linearization operator is non-self-adjoint, thus the
Newton-correction is obtained from solving the normal equation
(\ref{L1Dun2}), where the adjoint operator of $\textbf{L}_{1}$ is
\[
\textbf{L}_1^A=\textbf{L}_1^T=\left[ \begin{array}{cc}  L_{11} & L_{21} \\
L_{12} & L_{22} \end{array}\right].   \nonumber
\]

For Eq. (\ref{e:soliton}), the preconditioner in conjugate-gradient
iterations for solving the normal equation (\ref{L1Dun2}) is taken
as
\[ \textbf{M}=\mbox{diag}\left((\nabla^2+c)^2, \, (\nabla^2+c)^2\right),  \nonumber
\]
where $c$ is a positive constant (which we take as $c=3$ in our
computations).

While the above numerical algorithm is developed for real functions
$(\psi_1, \psi_2)$, during computer implementation, it is more
time-efficient to recombine $(\psi_1, \psi_2)$ into a complex
function $\psi$, so that the derivatives of $(\psi_1, \psi_2)$ can
be obtained simultaneously from $\psi$ by the fast Fourier
transform. Correspondingly, linear operators $\textbf{L}_{1}$ and
$\textbf{L}_{1}^A$ acting on real vector functions are combined into
scalar complex operations as well. Due to this recombination, the
code also becomes more compact.

Below we provide a sample Matlab code, where the asymmetric soliton
in Example 1 at $\mu=2.4$ is computed (see Fig.~2, at point `c'). On
a Desktop PC (Dell Optiplex 990 with CPU speed 3.3GHz), this code
takes 192 conjugate-gradient iterations and under 1.5 seconds to
finish with solution accuracy below $10^{-12}$.

\vspace{0.5cm}\noindent {\bf Matlab Code}
\begin{verbatim}
% In this code, U is the complex function psi
Lx=30; Ly=30; N=256;
errormax=1e-12; errorCG=1e-2; c=3;
x=-Lx/2:Lx/N:Lx/2-Lx/N;
y=-Ly/2:Ly/N:Ly/2-Ly/N;
kx=[0:N/2-1  -N/2:-1]*2*pi/Lx;
ky=[0:N/2-1  -N/2:-1]*2*pi/Ly;
[X,Y]=meshgrid(x,y); [KX,KY]=meshgrid(kx,ky);
K2=KX.^2+KY.^2; fftM=(c+K2).^2;
g=0.3*(exp(-(X+1.2).^2)+exp(-(X-1.2).^2));
gx=-0.6*((X+1.2).*exp(-(X+1.2).^2)+ ...
         (X-1.2).*exp(-(X-1.2).^2));
V=g.*g+10*g+i*gx;
sigma=1; mu=2.4;
U=1.2*exp(-(X-1.2).^2/2).*exp(-Y.^2/5);

tic
ncg=0;
while 1
  F=V+sigma*abs(U.*U)-mu; G=conj(F);
  L0U=ifft2(-K2.*fft2(U))+F.*U;
  errorU=max(max(abs(L0U)))/max(max(abs(U)))
  if errorU < errormax
     break
  end
  L1= @(W) ifft2(-K2.*fft2(W))+F.*W+ ...
           sigma*2*U.*real(conj(U).*W);
  L1A=@(W) ifft2(-K2.*fft2(W))+G.*W+ ...
           sigma*2*U.*real(conj(U).*W);
  DU=0*U;
  R=-L1A(L0U);
  MinvR=ifft2(fft2(R)./fftM);
  R2new=sum(sum(conj(R).*MinvR));
  R20=R2new;
  P=MinvR;
  while(R2new > R20*errorCG^2)
    L1P=L1(P); LP=L1A(L1P);
    a=R2new/sum(sum(real(conj(P).*LP)));
    DU=DU+a*P;
    R=R-a*LP; MinvR=ifft2(fft2(R)./fftM);
    R2old=R2new;
    R2new=sum(sum(real(conj(R).*MinvR)));
    b=R2new/R2old;
    P=MinvR+b*P;
    ncg=ncg+1;
  end
  U=U+DU;
end
ncg
toc
imagesc(x,y,abs(U)); colorbar; title('|\psi|')
\end{verbatim}

\end{document}